\documentstyle[12pt]{article}
\begin{document}
\thispagestyle{empty}
\begin{flushleft}
\end{flushleft}
\begin{center}
DROPLET-LIKE SOLUTIONS TO THE NONLINEAR SCALAR FIELD EQUATIONS IN THE
ROBERTSON-WALKER SPACE-TIME AND THEIR STABILITY \\
Yu.\ P.\ Rybakov$^{\star}$,\ B.\ Saha$^{\dagger}$,\ G.\ N.\ 
Shikin$^{\star}$\\ 
$^{\star}$ Department of Theoretical Physics, \\
Russian Peoples' Friendship University, Moscow \\ 
Miklukho Maklay St., 6, Moscow\,\, 117198, Russia \\
e-mail: rybakov@udn.msk.su\\
$^{\dagger}$ Bogoliubov Laboratory of Theoretical Physics,\\
Joint Institute for Nuclear Research, \\
141980\, Dubna, Moscow region, Russia\\
e-mail: saha@theor.jinrc.dubna.su 
\end{center} 
Exact droplet-like solutions to the nonlinear 
scalar field equations have been obtained in the Robertson-Walker 
space-time and their linearized stability has been proved.  \newpage 
\section{Introduction} 
\setcounter{equation}{0}

   When the general theory of relativity (GTR) and quantum theory 
of field were  developed,  an  interest  to  study  the  role  of 
gravitational interaction in elementary particle  physics  arose. 
On this context, to obtain and study the particle-like  solutions 
to the  consistent  systems  of  wave  and  gravitational  fields 
present a major interest.
  To  obtain  and  study  the  properties  of  regular  localized 
solutions to the nonlinear classical field equations (soliton- or 
particle-like solutions) is connected with the hope to develop  a 
divergence-free theory of elementary particle, which in its  turn 
would  describe  the  complex  spatial  structure  of   particle, 
observed experimentally. In doing so one should keep in mind that 
the  nonlinear  generalization  of  field  theory  is   necessary 
irrespective of the question of divergence as  the  consideration 
of interaction between the fields inevitably leads to the  advent 
of  nonlinear  terms  in  the  field   equations.   Consequently, 
nonlinearity should be considered not only as one of the ways  to 
eliminate difficulties of theory,  but  also  the  reflection  of 
objective properties of field. As it is noticed by N.\ N.\ Bogoluibov 
and  D.\ B.\ Shirkov [1], the  complete  description  of   elementary 
particles with all their physical characteristics (say,  magnetic 
momentum) can give only the interacting field theory. So one  can 
say that individual free (linear) fields  present  themselves  as 
the basis  to  describe  these  particles  in  the  framework  of 
interacting field theory. As elementary  particle  is  a  quantum 
object, so the attempts to develop a classical model of  particle 
remain   preliminary   but   necessary   stage   of   study   for 
transformation to quantum theory.

In this paper the system of interacting scalar and electromagnetic fields 
are being considered in the Robertson-Walker Universe with the metric [2]:
\begin{equation}
ds^2\,=\, dt^2\,-\, R^2(t)\biggl[\frac{dr^2}{1\,-\,kr^2}\, +\, 
r^2\,d\theta^2\,+\, r^2\,\mbox{sin}^2\theta\, d\phi^2 \biggr], 
\end{equation}
where $R(t)$ defines the size of the Universe, and $k$ 
takes the values $0$ and $\pm 1$. Droplet: it is some kind of soliton-like 
solutions to the field equations possessing sharp boundary. 
Similar solution was first obtained by Werle [3]. Further, a series of 
work was done where the solutions with sharp boundary to the nonlinear 
field equations were being found and studied in external gravitational 
field as well as in the selfconsistent one [4-10]. Present paper 
generalizes the partial results obtained by the authors earlier. Moreover 
here the question of stability is considered which presents a growing 
interest.

\section{Fundamental equations and their solutions} 
\setcounter{equation}{0}

We will choose the Lagrangian of interacting scalar $(\varphi)$ and 
electromagnetic $({\cal F}_{\alpha \beta})$ fields in the form [4]:  
\begin{equation}
{\cal L} =(1/2) \varphi_{,\alpha}\, \varphi^{,\alpha} -(1/4){\cal 
F}_{\alpha \beta}\, {\cal F}^{\alpha \beta}\, \Psi(\varphi), 
\end{equation} 
where the function $ \Psi(\varphi) = 1 + \kappa\, \Phi(\varphi)$  
characterizes the interaction ($ \Psi(\varphi) = 1$ corresponds to the 
system of free fields). We will seek the static spherically symmetric 
solutions assuming that the scalar field
$\varphi$ is the function of $r$ only, and the vector field 
${\cal A}_\mu$ possesses one component ${\cal A}_0 
(r)$, i.e.  $$ \varphi\, =\, \varphi (r), \qquad {\cal A}_\mu \,=\, 
\delta_{\mu}^{0} {\cal A}_0 (r)\,=\, \delta_{\mu}^{0} {\cal A}(r).  $$ 
It means that $({\cal F}_{\alpha \beta})$ also possesses one component 
i.e.  $$ {\cal F}_{\alpha \beta}\,=\, 
(\delta_{\alpha}^{0}\, \delta_{\beta}^{1}\,-\, \delta_{\beta}^{0}\, 
\delta_{\alpha}^{1})\, {\cal F}_{01}(r)\,=\, {\cal A}^{\prime}(r), $$ 
where $\prime$ denotes differentiation with respect to $r$.

The equations to scalar and electromagnetic fields write:
\begin{equation}
\partial_{\nu}\,(\sqrt{-g}\, g^{\mu\nu}\, \varphi_{,\mu})\, + 
(\sqrt{-g}/2)\,{\cal F}_{\alpha \beta}\, {\cal F}^{\alpha \beta}\, 
\Psi_{\varphi}(\varphi)\,=\, 0, \quad \Psi_{\varphi}(\varphi) \,=\, 
\partial \Psi/ \partial \varphi, \end{equation} \begin{equation} 
\partial_{\nu}\,(\sqrt{-g}\,{\cal F}^{\mu \nu} \,\Psi(\varphi)) \,=\,0. 
\end{equation}

In accordance with the assumption, made above, the equation (2.3) is 
easily integrated at $r\,>\, 0$:  \begin{equation} {\cal F}^{01}(r)\,=\, 
q\, P(\varphi)/\sqrt{-g'}\,=\, q\, P(\varphi)\sqrt{1\,-\,k\,r^2}/R^3\,r^2, 
\end{equation} where $q\,=\,$const, \quad $P(\varphi)\,=\, 
1/\Psi(\varphi)$ and $-g'\,=\,-g/\mbox{sin}^2\theta\,=\, 
\frac{R^6\,r^4}{(1\,-\,k\,r^2)}$.

The equation (2.2) for $\varphi(r)$ in this case metamorphoses to the 
equation with "induced nonlinearity" [5]:  \begin{equation} 
\sqrt{-g'}\biggl(\sqrt{-g'}\, 
g^{11}\,\varphi^{\prime}\biggr)^{\prime}\,=\, q^2\, g_{11}\, P_{\varphi}. 
\end{equation} 
Let us make the following assuption of {\it the cosmological character of 
time} in Robertson-Walker Universe. Let us suppose that the 
cosmological time scale is {\bf much greater} than the usual time scale. 
In other words in the case considered $R(t)$ can be interpreted as a 
constant. Then it is also easy to find the first integral and solution in 
quadrature for the equation (2.5):  \begin{equation} \varphi^{\prime} 
\,=\, -\sqrt{2}\,q\,\sqrt{P\,+C}/R\,r^2\,\sqrt{1\,-\,k\,r^2}, \quad C\,=\, 
\mbox{const}, \end{equation}
\begin{equation}
\int d\varphi/\sqrt{P\,+\,C}\,=\, \sqrt{1\,-\, k\,r^2}/R\,r \,+\, C_3.
\end{equation}
The regularity condition of $T_{0}^{0}$ at the center leads to the fact 
that $C\,=\,0$.  Choosing $P(\varphi)$ in the form \begin{equation} 
P(\varphi)\,=\,1/\Psi(\varphi)\, =\, J^{2-4/\sigma}\biggl(1\, -\, 
J^{2/\sigma}\biggr)^2, \end{equation}
where $J\,= \, \lambda\,\varphi, \quad \sigma\,=\, 2n+1, \quad n\,= \, 
1,\, 2 \cdots$, for $\varphi (r)$ one gets: 
\begin{equation} \varphi (r)\, =\, \frac{1}{\lambda} \biggl[1\, -\, 
\mbox{exp}\, \biggl(-\frac{2\sqrt{2}\,q\, \lambda}{R\, 
\sigma}\sqrt{1/r^2\, - k} + C_3\biggr)\biggr]^{\sigma/2}.
\end{equation} It is obvious that at 
$ r\to 0 \quad \varphi(0) \to 1/\lambda,$ 
and beginning with some 
$$ 
r\,=\, r_c\,=\, 2\sqrt{2}q\,\lambda / \sqrt{(R^2\, \sigma^2\, 
C_{3}^{2}\,+\, 8\, k\,q^2\,\lambda^2)},$$
$\varphi(r)$ becomes totally imaginary as in this case the square bracket 
possesses negative value. As far as we are dealing with a real scalar 
field, $\varphi (r)$ at $r\,>\, r_c$ becomes non-physical. So without 
losing the generality we may write that at $r\,\to r_c,
\quad \varphi(r_c)\to 0.$
(An illustration of the inverse interaction function $P(\varphi)$ and the 
scalar field obtained is given in Figure 1 and Figure 2.)

Let us write the energy-momentum tensor for the interacting fields: 
\begin{equation} 
T_{\mu}^{\nu}\,=\, \varphi_{,\mu}\, \varphi^{,\nu}\, -\,{\cal F}_{\mu 
\beta}\, {\cal F}^{\nu \beta}\, \Psi(\varphi)\,-\, 
\delta_{\mu}^{\nu}\,{\cal L}. \end{equation} From (2.10) we find 
the density of field energy of the system: 
\begin{equation} T_{0}^{0}\,=\, \frac{3}{2}\, 
\frac{q^2\,P}{R^4\,r^4} \end{equation} and total energy 
\begin{equation} E_f\,=\, \int T_{0}^{0}\,\sqrt{-^3g}\, d^3{\bf x}\,=\, 
\frac{3\sqrt{2}\,q\,\pi}{2\lambda\,(\sigma-1)}.  \end{equation} Thus, we 
came to the conclusion that energy density $T_{0}^{0}$ and total energy of 
the configurations obtained do not depend on the conventional values of 
the parameter $k\,=\, 0,\, \pm 1$.

\section{Stability problem}
\setcounter{equation}{0}

To study the stability of the configurations obtained we will write the 
linearized equations for the radial perturbations of scalar field. 
Assuming that 
\begin{equation} \varphi (r,\, t)\,=\, \varphi (r) + \xi(r,\,t), \quad \xi 
\ll \varphi, \end{equation} from (2.2) in view of (2.5) we get the 
equation for $\xi(r,\, t):$ \begin{equation} \ddot {\xi} \,+\, 3\, 
\frac{\dot R}{R}\, \dot \xi\,-\, \frac{1\,-\, k\, r^2}{R^2}\, \xi^{\prime 
\prime}\,-\, \frac{2\,-\, 3\, k\,r^2}{r\, R^2} \, \xi^{\prime}\,+\, 
\frac{q^2\, P_{\varphi \varphi}}{R^4\, r^4}\, \xi\,=\, 0.  \end{equation} 
As far as according to the assumption the external gravitational field is 
cosmological one, we can consider that $R(t)$ is a slowly varying 
time-function:  $\dot {R}(t) \approx 0$.  Assuming that \begin{equation} 
\xi(r,\,t)\, \approx\, v(r)\, \mbox{exp}(-i\, \Omega\, t), \quad 
\Omega\,=\, \omega/\,R, \end{equation} from (3.2) we obtain 
\begin{equation} (1\,-\, k\,r^2)\, v^{\prime \prime}\,+\, (2/\,r\,-\, 3\, 
k\,r)\, v^{\prime}\,+\,(\omega^2 \,-\, \frac{q^2\, P_{\varphi 
\varphi}}{R^4\, r^4})\, v\,=\, 0.  \end{equation} Let us first consider 
the case when $k\,=\, +1$. Then substituting $v(r)\,=\,y(x)$, where
$x\,=\,1\,-\,1/r^2,$ from (3.4) we get the equation \begin{equation} 
4\,x\,y_{xx}\,+\,2\,y_x\,+\, \biggl(\frac{\omega^2}{(1\,-\,x)^2} \,-\, 
\frac{q^2\, P_{\varphi \varphi}}{R^4}\biggr)\, y\,=\, 0, \end{equation} 
which for $y(x)\,=\,u(z), \quad x\,=\,z^2,$ takes the form
\begin{equation} u_{zz}\,+\, \biggl(\frac{\omega^2}{(1\,-\,z^2)^2} \,-\, 
\frac{q^2\, P_{\varphi \varphi}}{R^4}\biggr)\, u\,=\, 0.  \end{equation} 
Further substitution $$\eta(\zeta)\,=\, u(z)/\,\sqrt{1\,-\,z^2}, \quad 
z\,=\,-\mbox{th}\,\zeta,$$ leads the equation (3.6) to the normal form 
of Liouville [11] \begin{equation} \eta_{\zeta\zeta}\,+\, 
\biggl(\omega^2\,-\,1\,-\, \frac{q^2\, P_{\varphi 
\varphi}}{R^4}\,\mbox{sech}^4\,\zeta\biggr)\, \eta\,=\, 0. \end{equation} 
In case of $k\,=\,-1$ the equation (3.4) can analogously be transferred 
to the form (3.7) doing the following substitutions: $x\,=\,1\,+\,1/r^2, 
\quad x\,=\,z^2$ and $z\,=\, \mbox{th}\,\zeta.$

At last in case of $k\,=\,0$ from (3.4) we get
\begin{equation}
W^{\prime\prime}\,+\, \biggl(\omega^2\,-\, 
\frac{q^2\, P_{\varphi \varphi}}{R^4\,r^4}\biggr)\, 
W\,=\, 0,  \end{equation} 
where $W(r)\,=\, r\cdot v(r).$

Using the form of $P_{\varphi\varphi}$ from (2.8), we come to the 
conclusion that for $\sigma\, \ge\,5$ the expressions of the potentials $$ 
V_{\pm}(\varphi)\,=\,1\,+\, \frac{q^2\, P_{\varphi \varphi}}{R^4}\, 
\mbox{sech}^4\,\zeta \quad \mbox{and} \quad V_0(\varphi)\,=\, \frac{q^2\, 
P_{\varphi \varphi}}{R^4\,r^4} $$ tends to $+\infty$ at $r\,\to\,0$  as 
well as at $r\,\to\, r_c\,=\, \frac{2\sqrt{2}q\,\lambda}{ \sqrt{(R^2\, 
\sigma^2\, C_{3}^{2}\,+\, 8\, k\,q^2\,\lambda^2)}}$. It means that for 
$\sigma\,\ge\,5$ for $P(\varphi)$ given by (2.8) the configuration 
obtained is stable for the class of perturbation, vanishing at 
$r\,=\,0$ and $r\,=\,r_c$.

In stability can be assured in general introducing the variable 
$$\zeta\,=\,-\,\int\limits^{r}\,\frac{dr}{\sqrt{-g'}\,g^{11}}\,=\, 
\frac{1}{R}\int\limits^{r}\,\frac{dr}{r^2\, \sqrt{1\,-\,k\,r^2}}$$
and rewriting the equation for perturbation in the form 
$$\frac{d^2\,\xi}{d\zeta^2}\,+\, (\Omega^2 
\,-\, q^2\,P_{\varphi\varphi}) \,\xi\,=\,0.$$ The equation 
mentioned possesses at $\Omega\,=\,0$ 
nonnegative solution 
$\xi\,=\,-\,d\varphi/\,d\zeta$, which according to the Sturm theorem 
corresponds to the absence of "coupled" state with 
$\Omega^2\,<\,0.$  

\section{Conclusion}
\setcounter{equation}{0}

   Thus, we obtain the object with sharp boundary,  described  by 
the regular function $\varphi(r)$. In  the  center  of  
the  system $r\,=\,0$ \quad $\varphi(0) \to 1/\lambda$, and at some 
critical value of radius $r\,=\,r_c$  function 
$\varphi(r)$ possesses trivial value.  The configuration obtained, 
possesses limited energy density and finite total energy. From (2.12) it 
is explicit that the expression for energy does not  contain $r$, defining 
the size of droplet.  It  means that  the droplets  of different linear 
sizes up to the soliton with $r_c\,\to\,\infty$ share one and the same 
total energy.  For different values of $k$, the 
field  function $\varphi(r)$ changes it's  form.  It is noteworthy to 
notice  that  at $r_c\,\to\,\infty$ for $k\,=\,0$ droplet 
transfers to usual solitonian solution, while in case of  
$k\,=\,\pm 1$ this type of transition remains 
absent.  It should also be emphasized that the values $k\,=\,\pm 1$ 
enforce the stability of the configurations obtained, 
which is obvious from the expressions of $V_0(\varphi)$ and 
$V_{\pm}(\varphi)$.

\end{document}